\documentstyle [epsf,epsfig,12pt]{article}

\def\beqa{\begin{eqnarray}}
\def\eqa{\end{eqnarray}}

\newcommand{\PO}{I\!\!P}
\newcommand{\RO}{I\!\!R}
\newcommand{\xpom}{x_{\PO}}

\def\beq{\begin{equation}}
\def\eq{\end{equation}}

\renewenvironment{thebibliography}[1]
          {\begin{list}{[\arabic{enumiv}]}
          {\usecounter{enumiv}\setlength{\parsep}{0pt}
\setlength{\leftmargin .75cm}{\rightmargin 0pt}
        \setlength{\itemsep}{8pt} \settowidth
        {\labelwidth}{#1.}\sloppy}}{\end{list}}
\begin{document}

\input epsf

\begin{titlepage}
\vspace*{-1.5cm}

\begin{center}
\baselineskip=13pt

{\hspace*{10cm} Saclay-SPP, DESY 98 
  }
\vspace{2cm}

{\Large \bf  A Parametrisation of the Inclusive Diffractive Cross 
Section at HERA\\}
\vskip2.5cm
{\Large Jochen Bartels}\\
\vskip1cm
{\it II. Institut f\"ur Theoretische Physik, Universit\"at Hamburg,
\\ Luruper Chaussee 149, D-22761 Hamburg}\\
\vskip1.5cm
{\Large Christophe Royon}\\
\vskip1cm
{\it Service de Physique des Particules, DAPNIA, CEA-Saclay, \\
F91191 Gif sur Yvette Cedex, France}\\
\vskip1.5cm
\end{center}
\vspace*{1.5cm}
\begin{abstract} 
A recently proposed parametrization for the deep inelastic diffractive cross 
section is used to describe the H1 94 data. We find two possible solutions,
and we discuss them is some detail.
\end{abstract}

\end{titlepage}
\mbox{}
{\bf 1.} Diffractive events at HERA ~\cite{F2DH194,zeus} are generally 
attributed to the 
exchange of the Pomeron between the virtual photon and the proton.
After the first observation of these events, attempts of
a theoretical description ~\cite{BGF} where based upon the
idea of Ingelman and Schlein ~\cite{IS}, and they viewed diffraction
in deep inelastic scattering as ``deep inelastic scattering
on a Pomeron``. As a striking result along these lines, H1 ~\cite{F2DH194}
reported on
a ``hard gluon component of the Pomeron`` (where ``hard`` refers to the 
behavior of the structure function near $\beta=1$). However, with increasing
statistics both ZEUS and H1 ~\cite{F2DH194,zeus} found evidence that in DIS diffraction the Pomeron
intercept tends to be higher than the soft hadronic Pomeron. This observation
suggested that the diffractive cross section at HERA contains both a soft 
and a hard component, and the latter one is stronger than initially expected.

A theoretical description of the inclusive diffractive cross section,
therefore, should contain both a nonperturbative Pomeron part and a 
perturbative contribution. As a possible strategy, one could 
start from that part of the final state where pQCD can be used safely,
and then attempt to find a smooth extrapolation into the nonperturbative
Regge region. As to the Pomeron, in the perturbative region it is most 
easily modelled by a two gluon-system, the gluon structure function of the 
proton. In the nonperturbative region, this model of a ``hard Pomeron`` 
should smoothly connect with the soft hadronic Pomeron.
Recently ~\cite{bartels} a parametrization of the diffractive 
cross section has been
proposed which follows this line of arguments, and it has been applied to
both ZEUS and H1 data. For the latter the kinematic region was restricted
to those values of $\beta$, $Q^2$, and $x_{\PO}$ for which also ZEUS data
points exist. As a striking result, it was found that the ZEUS data
allow for a unique and rather succesful description whereas for the H1
data two different fits of this model were found.

In the present paper we apply the parametrization of ~\cite{bartels} to the
complete H1 94 data of the diffractive cross section. We confirm the existence
of two different solutions and we study their properties in some detail.

Let us first briefly recapitulate the parametrization.
It consists of 3 terms each of which stands for a particular diffractive
final state:
\begin{eqnarray}
&~& F_2^{D(3),I} = A \left( \frac{x_0}{x_{\PO}}
\right)^{n_2} \beta (1-\beta) \\
&~& F_2^{D(3),II} = B \left( \frac{x_0}{x_{\PO}}
\right)^{n_2}  \alpha_S 
\left( ln (\frac{Q^2}{Q_0^2} +1) \right) (1-\beta)^{\gamma} \\
&~& F_2^{D(3),III} = C \left( \frac{x_0}{x_{\PO}}
\right)^{n_4}  \left( \frac{Q_0^2}{Q^2} \right) 
\left( ln (\frac{Q^2}{4 Q_0^2 \beta} +1.75) \right)^2 
\beta^3 (1-2 \beta)^{2}
\end{eqnarray}
where
\begin{eqnarray}
n_{2,4}= n^0_{2,4} + n^1_{2,4} ln \left[ ln \left( \frac{Q^2}{Q_0^2} \right) 
+1 \right] 
\end{eqnarray}
The first term describes the diffractive production of a $q \bar{q}$ pair from
a transversely polarized photon, the second one the production of 
a diffractive $q \bar{q} g$ system, and the third one the production of a
$q \bar{q}$ component from a longitudinally polarized photon.
The ansatz for the $\beta$-dependence is motivated by rather general features 
of QCD-parton model calculations: At small $\beta$ (large diffractive masses
$M_X$) the spin 1/2 (quark) exchange in the $q\bar{q}$ production (1) 
leads to a
behavior $\sim \beta$, whereas the spin 1 (gluon) exchange in (2) corresponds 
to $\beta^0$. For large $\beta$ (small diffractive masses) in (1) and (3) 
calculations show ~\cite{calc,BLW} that pertubative QCD becomes reliable, and they lead to 
$1-\beta$ and $(1-\beta)^0$ in (1) and (3), resp. For the $q\bar{q}g$
term in (2) the situation is slightly more complicated: a QCD calculation
~\cite{Wu}
leads to a $(1-\beta)^3$ behavior. But in view of the previous H1 result
which suggests a hard gluon distribution inside the
Pomeron and, therefore, also a large gluon contribution in $F_2^D$ near
$\beta=1$, we will leave the exponent
$\gamma$ as a free parameter. Comparing the behavior in $\beta$
near $1$ of the three terms we notice that the longitudinal term (3) dominates.

Turning to the $Q^2$ dependence, the first two terms are leading twist
whereas the last one belongs to twist 4. Combining this with the
$\beta$ dependence discussed before we see that the longitudinal 
term is important only near $\beta=1$. The $\ln Q^2$-terms follow from
QCD-calcutions, and they indicate the beginning of the QCD $Q^2$-evolution.
In the transverse case ((1) and (2)) the evolution starts only after the
radiation of at least one gluon, whereas for the higher twist longitudinal 
case the first logarithm appears already in the quark loop. The longitudinal
$q\bar{q}g$ production contributes to leading twist, but it does not
have the $\log Q^2$ enhancement and therefore is small compared to the 
transverse term (2). It should be clear that the parametrization (1) - (3)
does not yet aim at describing the $Q^2$ evolution over a large $Q^2$ region.
At this early stage it is designed only for the small-x HERA region 
where $Q^2$ stays below 
$100$ GeV$^2$. The use in the large-$Q^2$ region will definitely require some
refinement of the parametrization.
Finally, the dependence on $x_{\PO}$ cannot be obtained
from perturbative QCD and therefore is left free. For the two transverse terms
one expects approximately the same $x_{\PO}$-behavior, not too far away
from the soft Pomeron. For the longitudinal term (3), on the other hand,
it has been shown ~\cite{BLW} that the $x_{\PO}$ dependence should be
approximately the same as for the square of the gluon structure function.
In the fit both $n_2$ and $n_4$ are allowed to have have a linear rise
in $\ln \ln Q^2$.

In contrast to the ZEUS diffraction data H1 data extend into the region of
not so large rapidity gaps where, in the Regge language, also the
exchange of secondary reggeons has to be taken into account. Within the spirit
of the parametrization proposed in ~\cite{bartels} such a term is very natural:
within pQCD it corresponds to the exchange of a $q\bar{q}$ system.  
Such a term has recently been studied in ~\cite{Scha}. In our present fit we
follow the H1 procedure of ~\cite{F2DH194} and use the following ansatz:
\begin{eqnarray}
F_2^{D(3),IV}(x_{\PO},\beta, Q^2)=
N f_{\RO / p} (x_{\RO}) F_2^{\RO} (\beta, Q^2)
\end{eqnarray}
where the reggeon flux is taken to follow a Regge behaviour with a linear
trajectory $\alpha_{\RO}(t)=\alpha_{\RO}(0)+\alpha^{'}_{\RO} t$ such
that:
\begin{eqnarray}
f_{\RO / p} (x_{\PO})= \int^{t_{min}}_{t_{cut}} \frac{e^{B_{\RO}t}}
{x_{\PO}^{2 \alpha_{\RO}(t) -1}} dt
\end{eqnarray}
where $|t_{min}|$ is the minimum kinematically allowed value of $|t|$
and $t_{cut}=-1$ GeV$^2$ is the limit of the measurement and the values
of $B_{\RO}$ and $\alpha^{'}_{\RO}$ are fixed with hadron-hadron 
data \cite{F2DH194}. $F_2^{\RO}$ is assumed to be proportional to the pion structure
function \cite{grv} \footnote{It was checked that assuming another pion 
structure function \cite{owens} for the reggeon does not modify the results
of the fits.}.
No interference is assumed between this reggeon exchange
and the Pomeron exchange of (1) - (3). The presence of (5) is a new
element in our fit.

In our fit the main parameters are the normalization constants A, B, C, N,
and the exponent $\gamma$ of the $\beta$-dependence near $\beta=1$.
Moreover, we allow the $x_{\PO}$ exponents to vary, i.e. we have the five  
parameters $n_2^0$, $n_2^1$, $n_4^0$, $n_4^1$, and the intercept 
$\alpha_{R}(0)$ of the secondary. Finally, we also leave $x_0$ as a free 
parameter.

{\bf 2.} Let us now turn to the results of our fit (table 1). Most important,
the data allow two different possible fits of the 1994 H1 data 
\footnote{To be more precise we mention that there are two more solutions: 
they differ from those in table 1 by their value of $x_0$. For the perturbative gluon 
solution we find $x_0=1.9~10^{-4}$, for the hard gluon solution 
$x_0=4.9~10^{-5}$. Their
$\chi^2$ values are substantially higher: $\chi^2=242.2$ and $\chi^2=223.8$
for the perturbative and for the hard gluon solution, resp. Therefore we will not discuss them
in our paper.}
The parameter which distinguishes the two solutions is $\gamma$:
in the first case we find $\gamma=8.24$, in the
second case $\gamma=0.27$. Both solutions have an
acceptable $\chi^2$ value, with a slight preference for the first one
(high gamma). If we perform the fit with statistical errors
only, the $\chi^2$ values per degree of freedom
of the two fits are 1.10/df and 1.25/df, resp.
An overall picture of the two solutions is given in Figures 1 and 2.

\begin{table}[tb]
\begin{scriptsize}
\begin{picture}(100,150)(0,0)
\put(5,80){\begin{tabular}{|c||c|c|c|} \hline
   &  standard &  $B=0$ & $C=0$ 
\\ \hline \hline
$\gamma$ & 8.24 $\pm$ 1.06 & -  & 9.43 $\pm$ 1.09 \\
\hline
$\alpha_{R}$ & 0.62 $\pm$ 0.02 & 0.88 $\pm$ 0.02 & 0.57 $\pm$ 0.08 \\
$N$ & 15.2 $\pm$ 3.0 & 3.6 $\pm$ 0.7 & 21.6 $\pm$ 11.5 \\
\hline
$A$ & 0.056 $\pm$ 0.005 & 0.027 $\pm$ 0.004 & 0.058 $\pm$ 0.006 \\
$B$ & 0.025 $\pm$ 0.003 & 0. (fixed) & 0.027 $\pm$ 0.004 \\
$C$ & 0.035 $\pm$ 0.016 & 0.074 $\pm$ 0.032 &  0. (fixed) \\
\hline
$n_2^0$ & 1.08 $\pm$ 0.04  & 1.13 $\pm$ 0.04 & 1.09 $\pm$ 0.03 \\
$n_2^1$ & 0.21 $\pm$ 0.03  & 0.31 $\pm$ 0.04 & 0.20 $\pm$ 0.03 \\
$n_4^0$ & 1.43 $\pm$ 0.08  & 1.37 $\pm$ 0.08 &  - \\
$n_4^1$ & 0.00 $\pm$ 0.05  & 0.00 $\pm$ 0.01 &  - \\
\hline 
$x_0$ & 0.40 $\pm$ 0.02 & 0.30 $\pm$ 0.02 & 0.41 $\pm$ 0.04 \\
\hline
$\chi^2$ & 184.6 & 216.0 & 284.7
\\ \hline
\end{tabular}}
\put(240,80){\begin{tabular}{|c||c|c|c|} \hline
   &  standard &  $B=0$ & $C=0$ 
\\ \hline \hline
$\gamma$ & 0.27 $\pm$ 0.18 & -  & 0.001 $\pm$ 0.024 \\
\hline
$\alpha_{R}$ & 0.80 $\pm$ 0.02 & 0.88 $\pm$ 0.02 & 0.76 $\pm$ 0.03 \\
$N$ & 5.7 $\pm$ 0.7 & 3.6 $\pm$ 0.7 & 7.2 $\pm$ 1.7 \\
\hline
$A$ & 0.059 $\pm$ 0.006 & 0.027 $\pm$ 0.004 & 0.045 $\pm$ 0.008 \\
$B$ & 0.014 $\pm$ 0.002 & 0. (fixed) & 0.027 $\pm$ 0.005 \\
$C$ & 0.104 $\pm$ 0.032 & 0.074 $\pm$ 0.032 &  0. (fixed) \\
\hline
$n_2^0$ & 1.20 $\pm$ 0.04  & 1.13 $\pm$ 0.04 & 1.33 $\pm$ 0.04 \\
$n_2^1$ & 0.20 $\pm$ 0.03  & 0.31 $\pm$ 0.04 & 0.09 $\pm$ 0.04 \\
$n_4^0$ & 1.41 $\pm$ 0.06  & 1.37 $\pm$ 0.08 &  - \\
$n_4^1$ & 0.00 $\pm$ 0.09  & 0.00 $\pm$ 0.01 &  - \\
\hline 
$x_0$ & 0.17 $\pm$ 0.02 & 0.30 $\pm$ 0.02 & 0.15 $\pm$ 0.01 \\
\hline
$\chi^2$ & 206.8 & 216.0 & 219.7
\\ \hline
\end{tabular}}
\end{picture}
\end{scriptsize}
\caption{Parameters obtained for the ``perturbative gluon'' (on the left) and 
``hard gluon''
(on the right) solutions of the
fit. The fits are performed with statistical and systematical
errors added in quadrature. The two last fits are preformed imposing
the $q \bar{q} g$ (B=0) or the longitudinal $q \bar{q} $ components
to be zero (C=0). If the fits are performed
with statistical errors only, the $\chi^2$ values for the perturbative gluon
and for the hard gluon solutions are 
236.0 (1.10/dof) and 269.1 (1.25/dof), resp.}
\label{t1.0}
\end{table}

\begin{figure}
 \begin{center}
   \mbox{\epsfig{figure=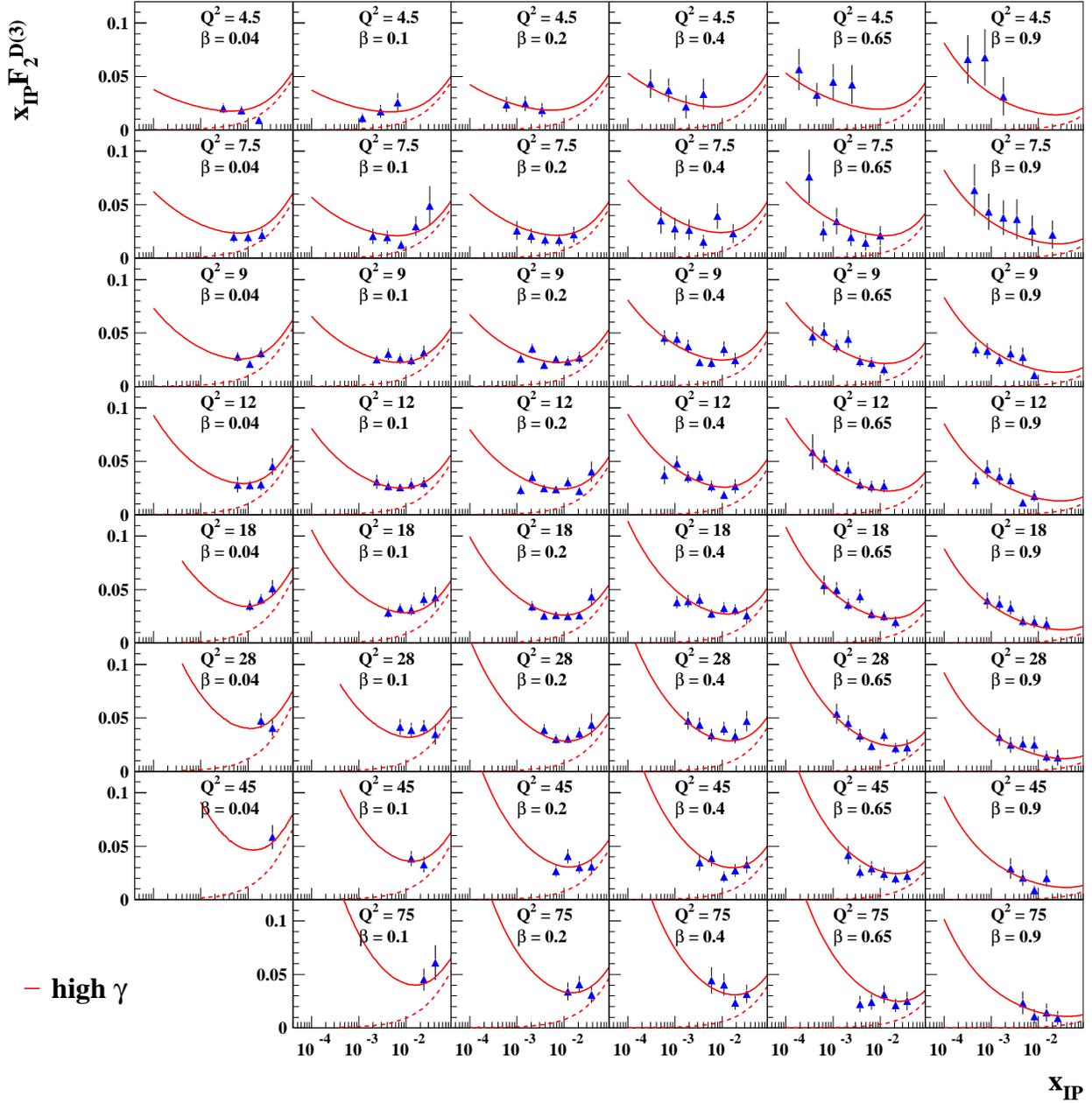,width=180mm}}
 \end{center}
 \caption{Result of the ``perturbative gluon'' fit. The triangles are the H1 1994 data.
 In full line is displayed the full result of the fit, and in dashed line, the
 reggeon component only which is important at low $\beta$ and high $\xpom$.}
 \label{high}
\end{figure}

\begin{figure}
 \begin{center}
   \mbox{\epsfig{figure=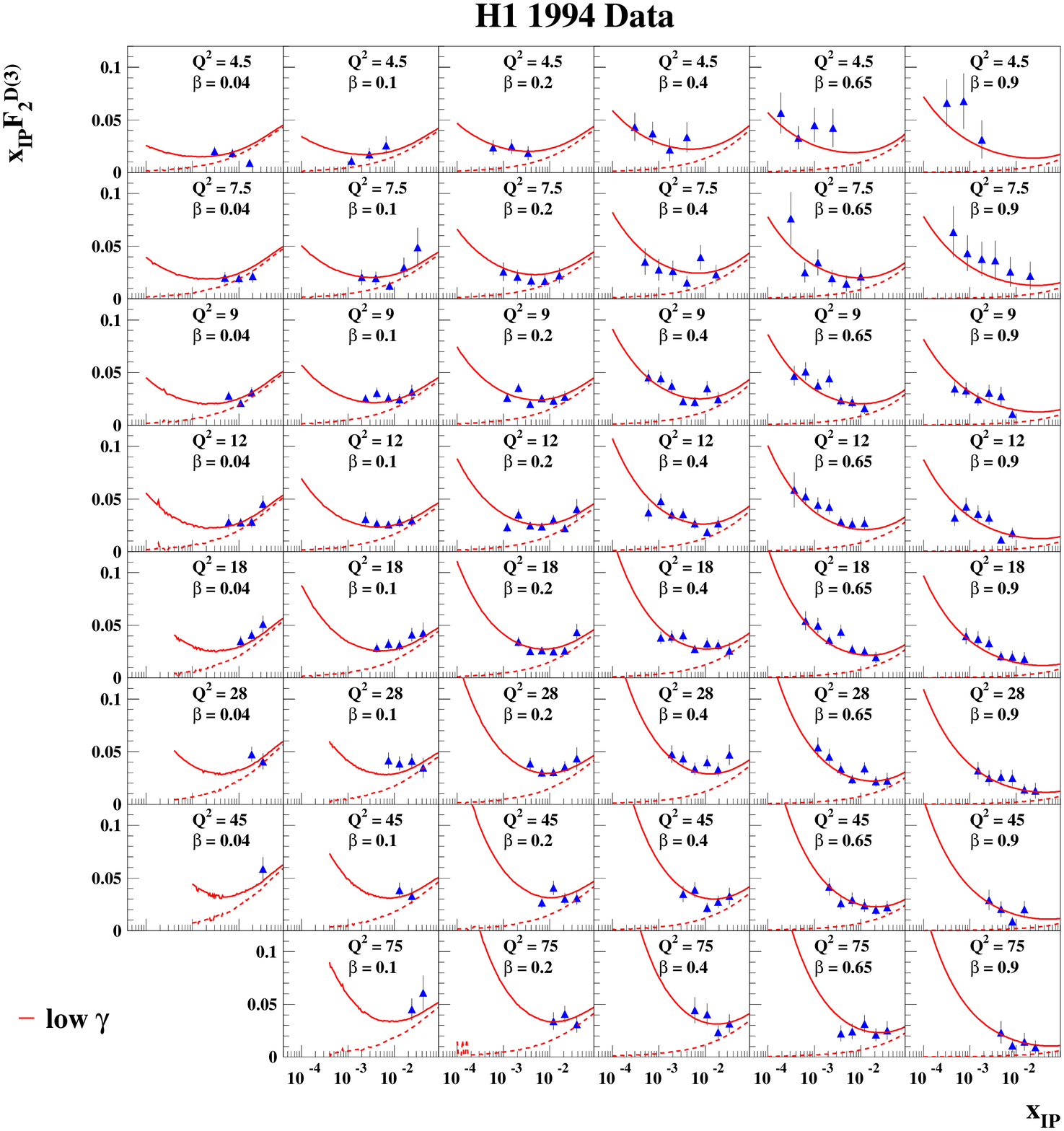,width=180mm}}
 \end{center}
 \caption{Result of the ``hard gluon'' fit. The triangles are the H1 1994 data.
 In full line is displayed the full result of the fit, and in dashed line, the
 reggeon component only which is important at low $\beta$ and high $\xpom$.
 This hard gluon fit is hardly distinguishable for the perturbative gluon one.}
 \label{low}
\end{figure}

The difference between the two solutions is most clearly illustrated
in Figs.3 and 4 where we plot the $\beta$-dependence of the different pieces
(1) - (4). Data points lie mostly in the central and the rhs columns
($x_{\PO}=0.001$ and $x_{\PO}=0.01$), but for illustration we also extrapolate
down to $x_{\PO}=0.0001$. The main difference lies in the behavior of the 
$q\bar{q}g$ component: for the first solution ($\gamma =8.24$) it contributes only
at small $\beta$ ($\beta<0.3$), whereas for the second one ($\gamma =0.27$)
it extends up to $\beta$ near one. Qualitatively this second solution
is close to the previous
H1 solution with the hard gluon: the fact that the gluon distribution 
obtained from the H1 QCD analysis peaks near $1$ leads to
a large gluon contribution to $F_2^D$ near $\beta=1$. In the following this
solution to our fit will 
therefore be referred to as the ``hard gluon'' solution.
The first solution, on the
other hand, one is closer to the perturbative prediction which leads to
$\gamma=3$. This solution will be called ``perturbative gluon'' solution in 
the following.
For both solutions one sees that the transverse $q\bar{q}$ term
has its maximum at medium $\beta$, and the longitudinal one contributes only near
$\beta=1$. For the perturbative gluon solution the transverse $q\bar{q}$ is somewhat
bigger than in the hard gluon case.

An important feature of both solutions is that none of the four components
(1) - (4) comes out to be particularly small. To illustrate the importance of
(2) and (3) we have repeated the fits (table 1) with putting either B or C 
equal to zero: in both cases the $\chi^2$ increases significantly. Let us notice
that only one solution remains if $B=0$ is required (both solutions in Table 1 are
identical). On the other hand, imposing $C=0$ in the hard gluon solution implies
a very low value of $\gamma$ ($\gamma=0.001$), and so an important
$q \bar{q} g$ term is present even at high $\beta$ and compensates the 
smallness of the $q \bar{q}$ term in this region.
The situation changes if we impose the same kinematic cuts as used in the
QCD fit of ~\cite{F2DH194}. Namely, by excluding  the data points at high
$y$ ($y>$0.45) and at low $M_X$ ($M_X <$2 GeV) this fit was restricted 
to the kinematic region where higher twist corrections are expected to
be small. Also, only $\beta$ and $Q^2$ bins with more than three measured
points were considered. Imposing the same cuts we have fitted our model to the
data: we still find the two solutions. However, as expected, the longitudinal
$q \bar{q}$ is now poorly constrained, and forcing the $C$ parameter to be 0 
in the fit (excluding the higher twist contributions) does not change the 
quality of the fit very
much (for the perturbative gluon solution the $\chi^2$ increases from 
130 to 133, whereas for the hard gluon it remains unchanged ($\chi^2=$144)).
The hard gluon solution now is even closer to the previous H1 solution 
in which no $q\bar{q}$ term was present.

The importance of the secondary reggeon term can be seen from Figs.1 and
2: in both cases its contribution becomes substantial only at low $\beta$ and
large $x_{\PO}$.

Scaling violations are illustrated in Figs.5 and 6. The $Q^2$ dependence 
of the longitudinal $q\bar{q}$ term (3) near $\beta=1$ is easily understood:
it decreases with increasing $Q^2$, but the decrease is much slower
than $1/Q^2$: this is due to the $(log Q^2)^2$ enhancement in (3). 
As to the other two terms (1) and (2), there is an additional 
$Q^2$-dependence coming from the exponent $n_2$: in both fits $n_2^1$
is not small, and in combination with the rather large values of the ratio
$x_0/x_{\PO}$ it leads to a substantial increase with $Q^2$. 
In Figs.5 and 6 this explains why also the transverse $q\bar{q}$ component 
rises with $Q^2$. For the $q\bar{q}g$ component, which according to (2) has
a logarithmic growth in $Q^2$,  we see a difference
between the two solutions: in the perturbative gluon solution it contributes only
at small $\beta$. Consequently, the rise in $Q^2$ of $F_2^{D}$
is stronger at small $\beta$ than at larger $\beta$ (e.g. $\beta=0.65$).
This effect is absent in the hard gluon solution, since here the gluon is 
present at all $\beta$.

In order to gain futher insight into the two solutions,
we have also performed fits
in a limited range in $\beta$ (or $M_X$).  The parameters obtained for the
perturbative gluon fit with cuts on $\beta$ in the data are given in Table 2.
We note that the parameters are quite stable except for $\gamma$ which has
the tendency to increase. However, the error on $\gamma$ is large, because
we cut explicitely the region at low $\beta$ (or large $M_X$) where the
$q \bar{q} g$ is significant. The hard gluon fit, on the other hand,
is much less stable:
already with the lowest cut at $\beta>0.1$ this solution disappears,
and only the perturbative gluon solution remains. Cuts on $M_X$ in the data 
lead to the same
results: excluding the high $M_X$ region, the perturbative gluon solution turns
out to be stable, whereas a cut $M_X \leq 10$ is already enough to eliminate 
the hard gluon solution. 

\begin{figure}
 \begin{center}
   \mbox{\epsfig{figure=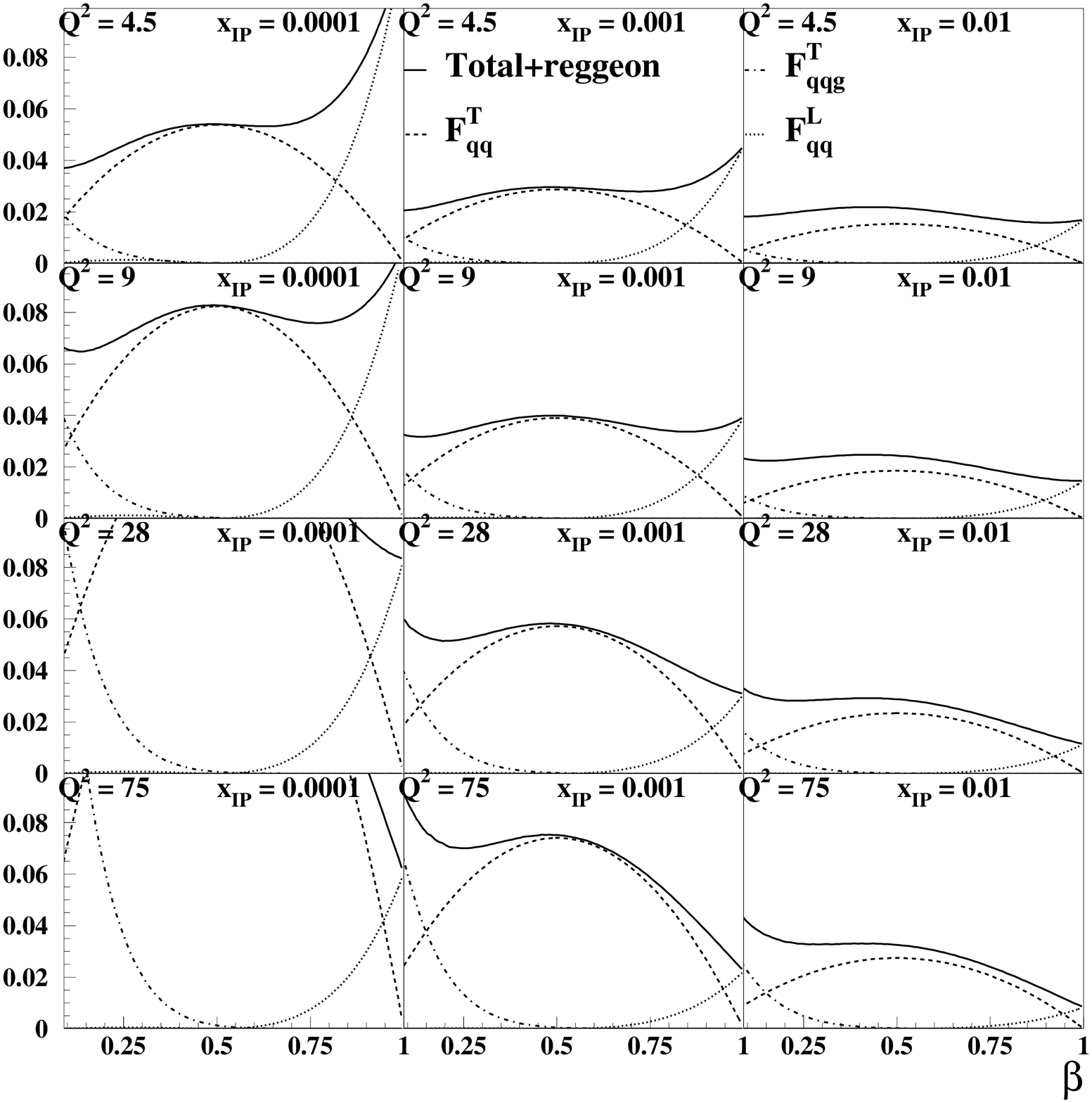,width=120mm}}
 \end{center}
 \caption{Display of the different components for the ``perturbative gluon'' fit.
 The transverse $q \bar{q}$ term belongs to the dashed line. The $q \bar{q} g$
 term (dashed-dotted line) is important at low $\beta$, and the longitudinal 
 $q \bar{q}$ term (dotted line) dominates at high $\beta$.}
 \label{highb}
\end{figure}

\begin{figure}
 \begin{center}
   \mbox{\epsfig{figure=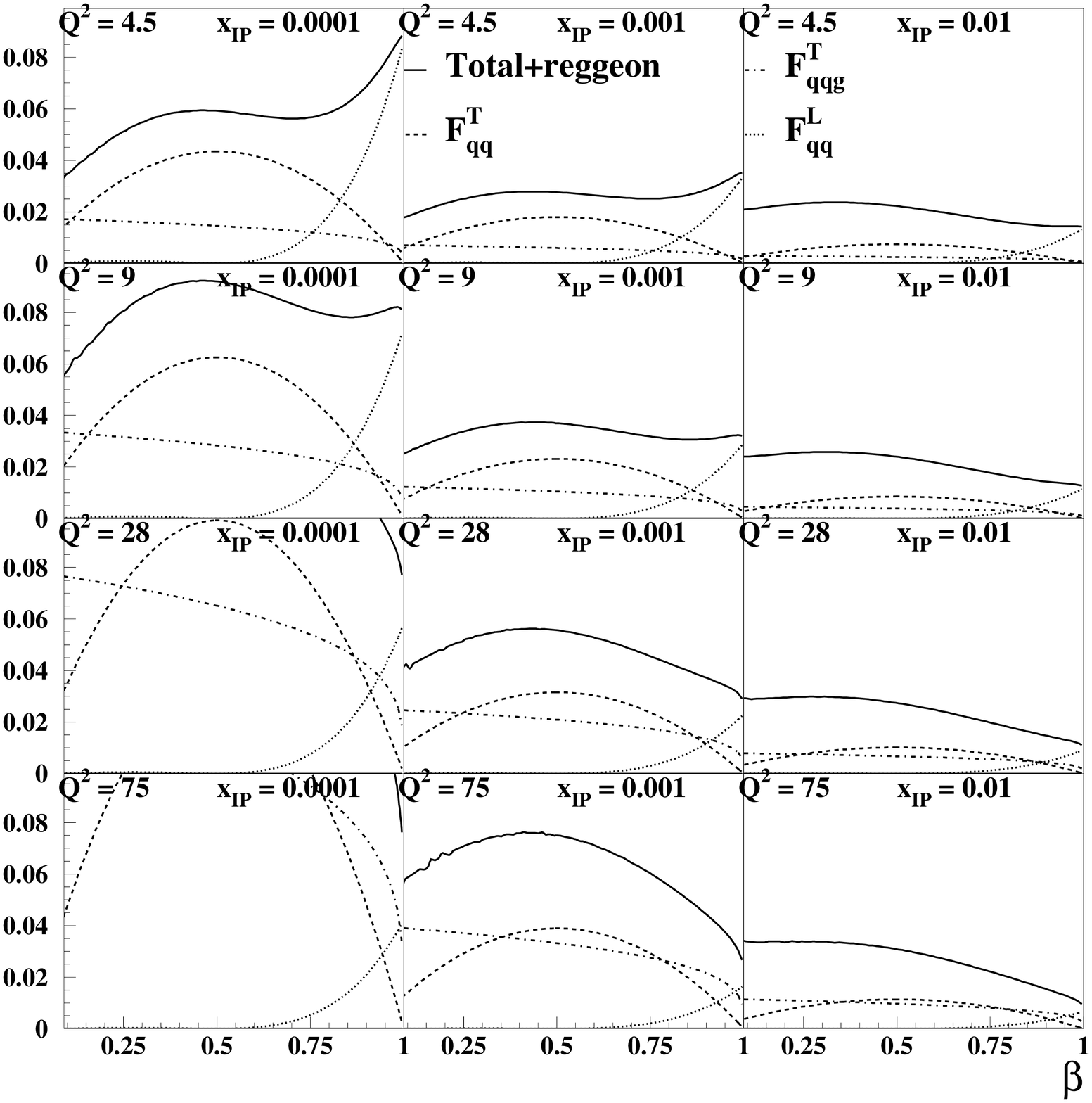,width=120mm}}
 \end{center}
 \caption{Display of the different components for the ``hard gluon'' fit.
 The transverse $q \bar{q}$ term belongs to the dashed line. The $q \bar{q} g$
 term (dashed-dotted line) is not much $\beta$-dependent, and the longitudinal 
 $q \bar{q}$ term (dotted line) dominates at high $\beta$.}
 \label{lowb}
\end{figure}

\begin{figure}[tb]
\begin{scriptsize}
\begin{picture}(100,350)(0,0)
\put(5,5){\mbox{\epsfig{figure=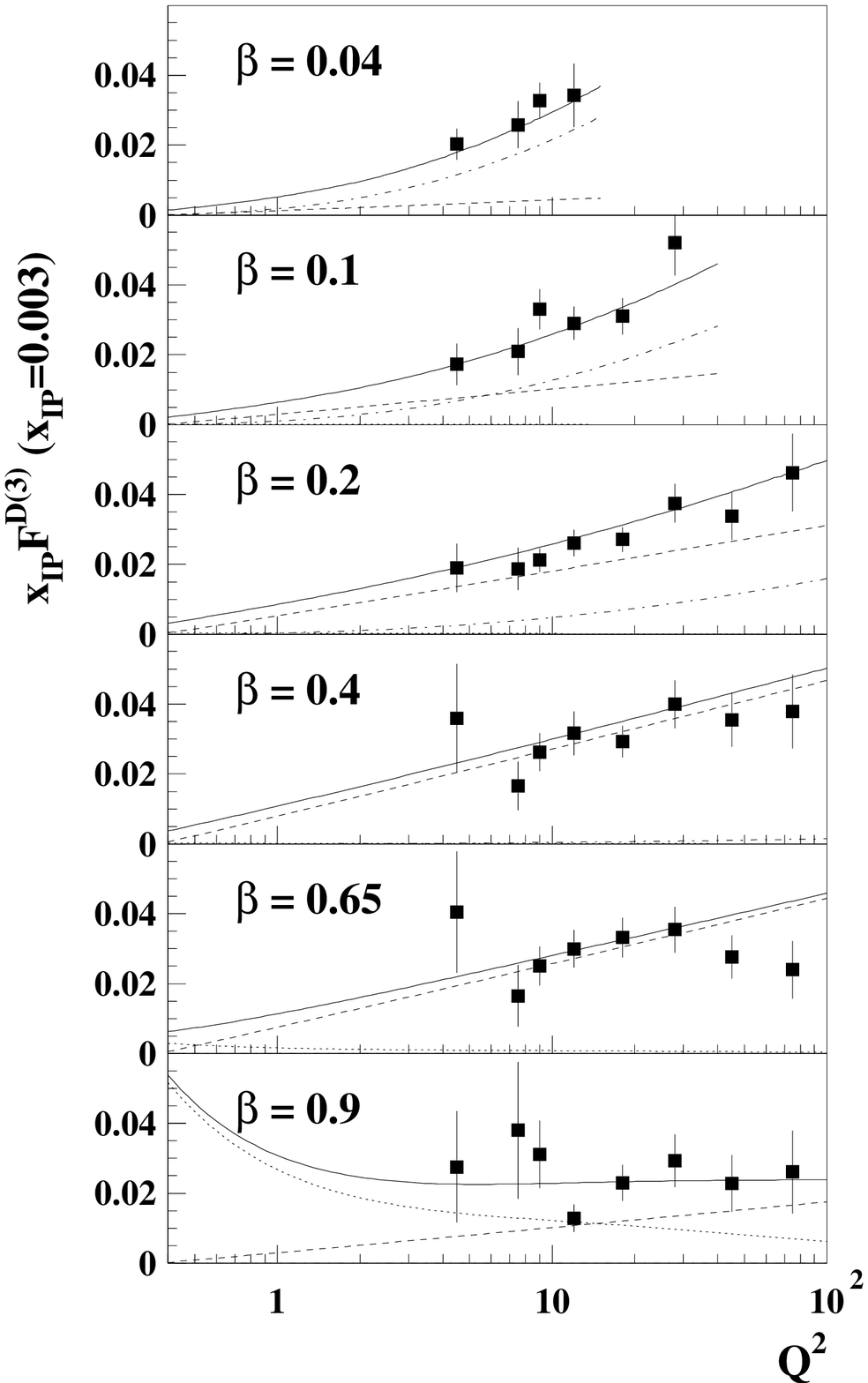,width=80mm}}}
\put(240,5){\mbox{\epsfig{figure=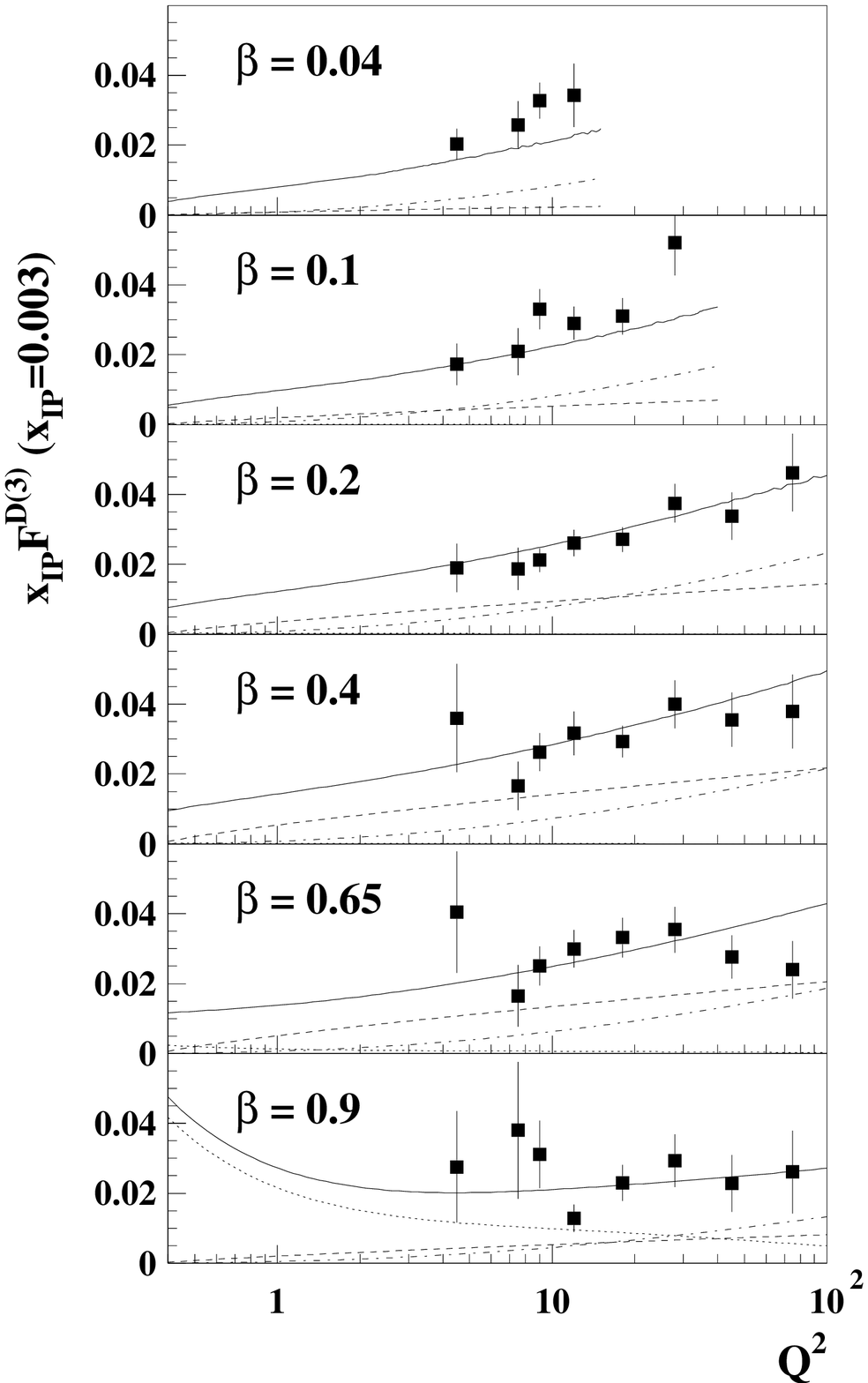,width=80mm}}}
\end{picture}
\end{scriptsize}
 \caption{Scaling violations in $\beta$ bins for a fixed value of $\xpom=0.003$
for the perturbative gluon (on the left) and hard gluon (on the right) solutions of the fit.
The transverse $q \bar{q}$, $q \bar{q} g$, and longitudinal $q \bar{q}$
components are respectively in dashed, dashed-dotted and 
dotted lines, respectively.
The full
line is the sum of all components, including the secondary reggeon.}
 \label{lowc}
\end{figure}

\begin{figure}[tb]
\begin{scriptsize}
\begin{picture}(100,350)(0,0)
\put(5,5){\mbox{\epsfig{figure=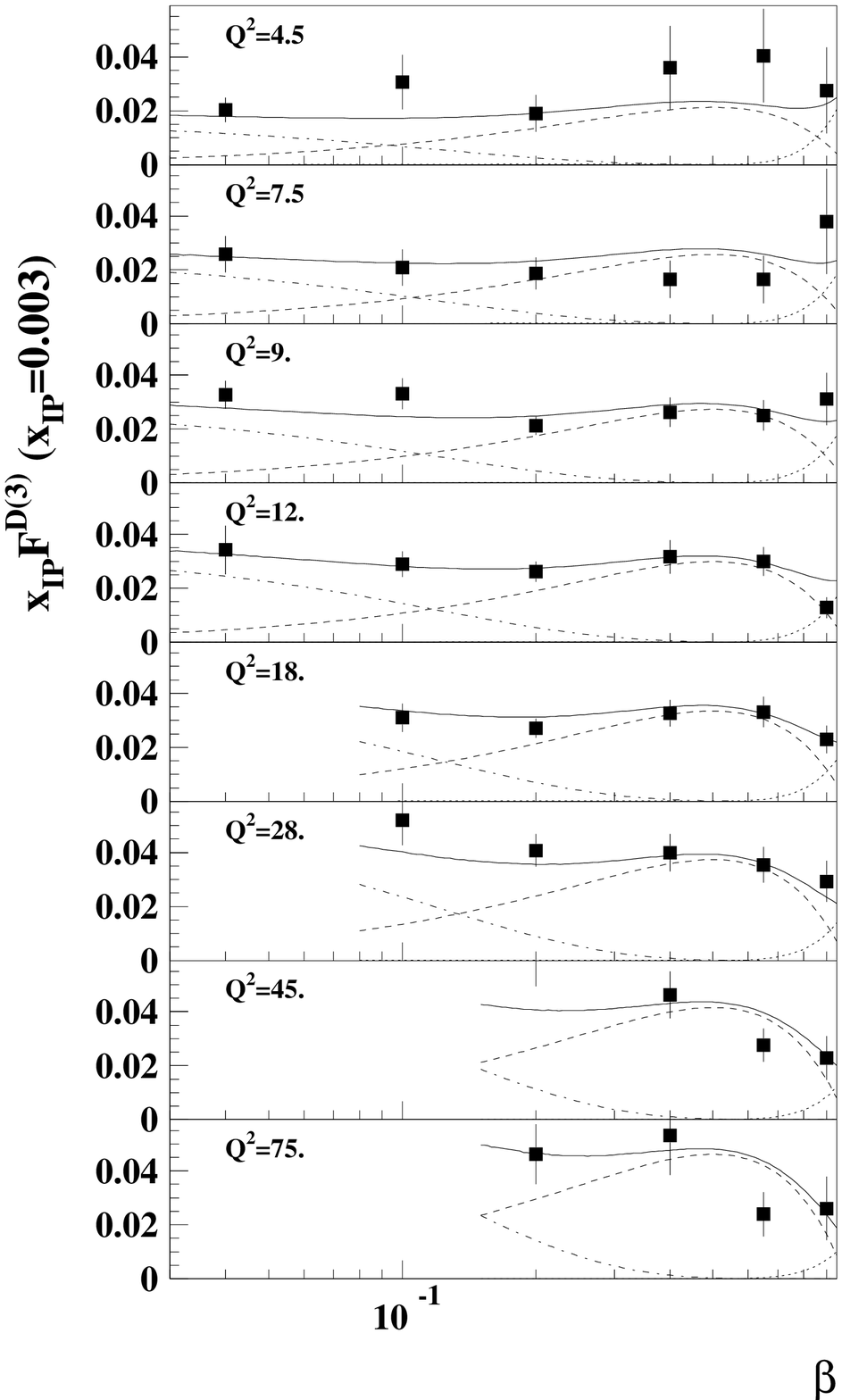,width=80mm}}}
\put(240,5){\mbox{\epsfig{figure=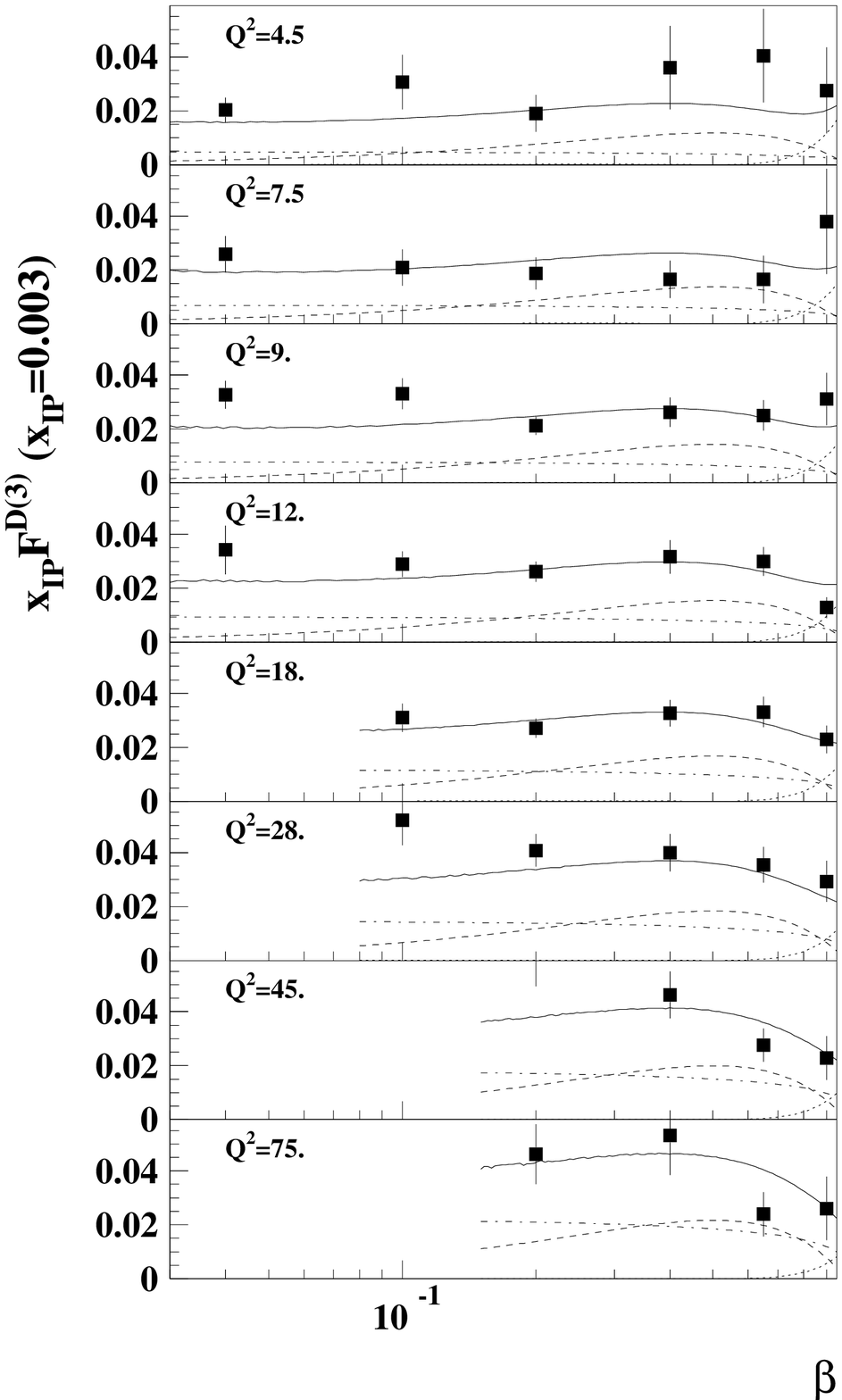,width=80mm}}}
\end{picture}
\end{scriptsize}
 \caption{$\beta$-dependence of $F_2^D$ in bins of $Q^2$ at a fixed value of $\xpom=0.003$
for the perturbative gluon (on the left) and hard gluon (on the right) 
solutions of the fit.
 The dashed, dotted and dashed-dotted lines represent
 the transverse $q \bar{q}$, 
 the longitudinal $q \bar{q}$ (important only at high $\beta$),
 and the $q \bar{q}g$ terms. The full
line is the sum of all components, including the secondary reggeon.}
 \label{lowc}
\end{figure}

\begin{table}
\begin{center}
\begin{tabular}{|c||c|c|c|c|c|} \hline
   &  standard &  $\beta >0.0.04$ & $\beta >0.1$ & $\beta >0.2$ &
$\beta >0.4$   
\\ \hline \hline
$\gamma$ & 8.24 $\pm$ 1.06 &  11.3 $\pm$ 2.2 & 15.2 $\pm$ 5.2 & 20. $\pm$ 18. &
 20. $\pm$ 17.  \\
\hline
$A$ & 0.056 $\pm$ 0.005 &   0.048 $\pm$ 0.012 & 0.048 $\pm$ 0.007 & 
0.051 $\pm$ 0.005 & 0.056 $\pm$ 0.005 \\
$B$ & 0.025 $\pm$ 0.003 & 0.030 $\pm$ 0.011 &  0.072 $\pm$ 0.080 & 
0.085 $\pm$ 1500. & 0.83 $\pm$ 9800. \\
$C$ & 0.035 $\pm$ 0.016 &  0.040 $\pm$ 0.018 & 0.037 $\pm$ 0.018 & 
0.036 $\pm$ 0.016 & 0.044 $\pm$ 0.018  \\
\hline
$\chi^2$ & 184.6 & 170.9 & 150.4 & 117.7 & 79.2
\\ \hline
\end{tabular}
\end{center}
\caption{Parameters obtained for the perturbative gluon solution of the
fit after different cuts on $\beta$ in the data.}
\label{t1.4}
\end{table}

{\bf 3.} Next we use our ansatz (1) - (4) for the ZEUS 94 data ~\cite{zeus}.
In agreement with ~\cite{bartels} we find only one solution, and the
$\gamma$-value turns out to be identical: $\gamma \sim 4.3$. 
Comparing this solution with the perturbative gluon 
solution of our H1 fit, we find that they are very similar,
both qualitatively and quantitavely. For instance, if in the H1 fit we fix
$\gamma$ at the ZEUS 
value $\gamma=4.3$, we get a $\chi^2$ of 197.8 which is not much bigger than 
the $\chi^2$ of the fit with all parameters left free ($\chi^2=186.7$). 
As an important result of our ZEUS fit, we note that the fit is not 
improved by adding the secondary reggeon: the reggeon normalisation is found
to be compatible with zero. This shows that
ZEUS data do not require any secondary contribution.

\begin{figure}
 \begin{center}
   \mbox{\epsfig{figure=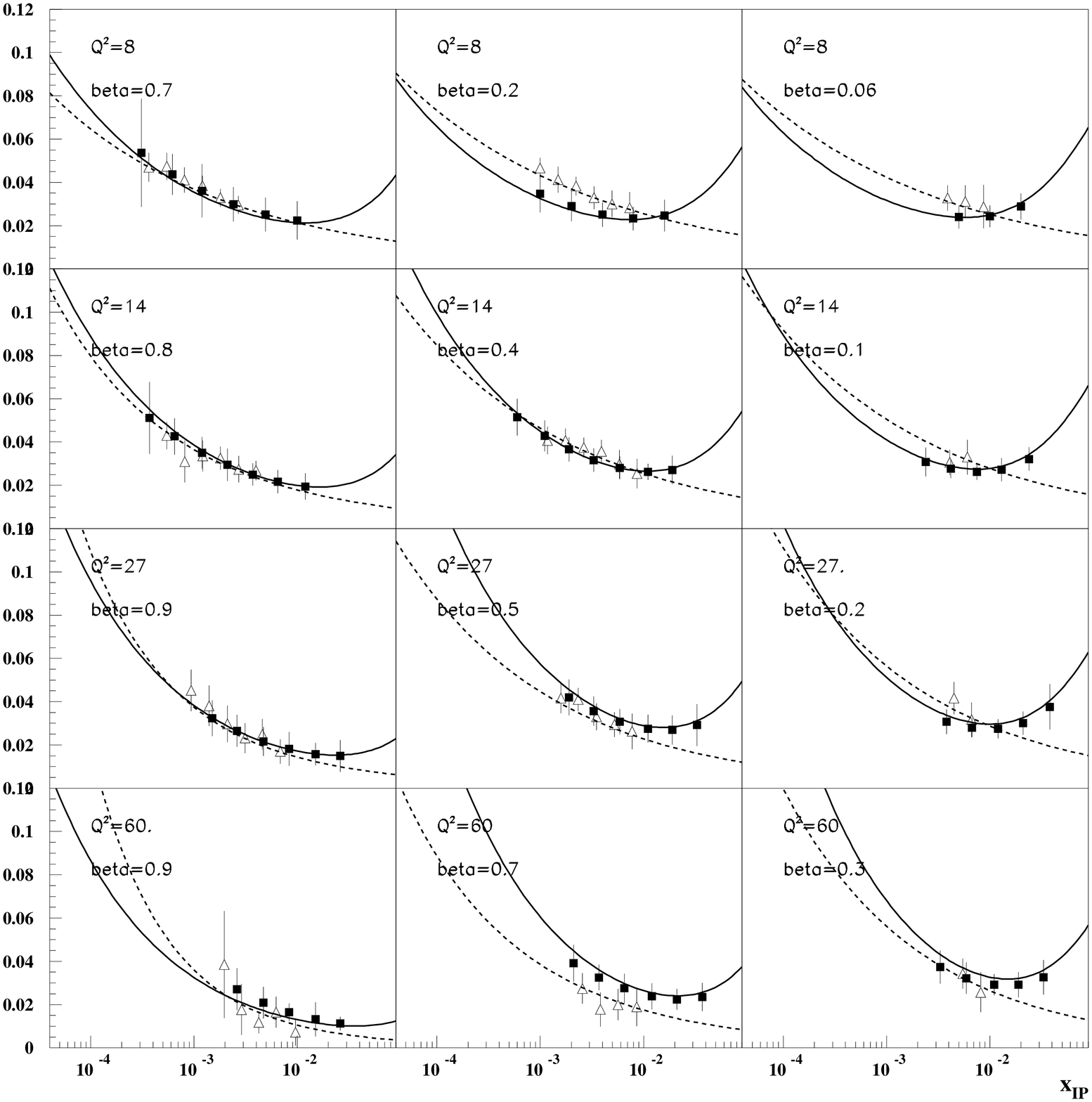,width=140mm}}
 \end{center}
 \caption{Comparison between H1 (squares) and ZEUS data (triangles). The full
 line represents the H1 perturbative gluon fit, and the dashed line our 
 ZEUS fit.
 We notice that the differences in the fits are precisely in the region where
 the data differ.}
 \label{zeus}
\end{figure}

To obtain a better understanding of the differences between the fits 
to the H1 data and to the ZEUS data ~\cite{zeus} , 
we have tried to compare the result of the H1 fit 
(the perturbative gluon solution) with the ZEUS data. The result is shown in 
Fig.\ref{zeus}. The triangles
represent the recently published ZEUS data, and the squares the H1 data. To be
able to compare directly H1 and ZEUS data, we had to shift the H1 data to the
$Q^2$ and $\beta$ ZEUS bins, using the perturbative gluon solution of the fit.
Let us notice that this comparison does not depend much on the way we perform
the extrapolation, as the $Q^2$ and $\beta$ bins from H1 and ZEUS are rather 
close. Further more, an extrapolation using a completely different fit based
on the dipole model give the same interpolated values \cite{dipole}. 
The full curve is then
the perturbative gluon fit result with H1 data. The hard gluon solution 
restricted to the ZEUS bins would
be indistinguishable on this plot. The dashed line denotes the result
of our fit to the ZEUS data. We notice that the fits differ precisely in 
that region
where one sees differences in the data, more precisely in the bins
$\beta=$0.2, $Q^2$=8, and, $\beta= $0.7, $Q^2=$60
\cite{dipole}. The differences in the
fit parameters are clearly due to differences in the data points. More precise
data in these regions will be thus very useful.

{\bf 4.} So far we have not discussed our results for the exponents $n_2$ and 
$n_4$. Since the errors for these parameters are not small, one may feel that
at this stage we should not attribute too much weight to them.
Nevertheless, in the following we describe a few fits in which we impose
constraints on $n_2$ and/or $n_4$. First we try to identify the $x_{\PO}$
dependence in (1) and (2) with the soft Pomeron, leaving $n_4$ as a free
parameter. We put $n_2=1.12$ (this differs from the soft Pomeron value
$2\alpha_{\PO}(0) - 1 =1.16$ by a small amount which is due to the integral
over the momentum transfer $t$) and $n_2^1=0$, i.e. we exclude 
any $Q^2$ variation. The results for both solutions are shown in 
table 3. The obtained $\chi^2$ are quite bad: 254.2 (291.2) to be 
compared with 206.8 (184.6) in table 1 for the 
perturbative (hard) gluon solutions. This makes it unlikely that  
the diffractive structure function data could be described by the soft pomeron.
In a variant of 
this test we have also treated $n_2^0$
as a free parameter (still keeping $n_2^1=0$). For the perturbative
gluon solution we find a reasonable fit 
($\chi^2=221.9$) for $n_2^0=1.38$, and for the hard gluon solution  
the fit gives $n_2^0=1.23$ with $\chi^2=239.6$. 

Next we test the possibility of identifying the $x_{\PO}$ dependence 
of the longitudinal term with the gluon density in the proton.
We take the gluon parametrisation from \cite{grv94} which is known
to give a good description of the proton structure function data 
measured at HERA and fit it to the form $x^{-\alpha +1}$.
We find 
\begin{eqnarray}
xG \sim x^{-\alpha +1}=x^{-0.18-0.15 \ln ( \ln (Q^2)+1)}
\end{eqnarray}
where the scale of $Q^2$ is 1 GeV$^2$.
This suggests to parametrize the exponent $n_4$ in the following way:
\begin{eqnarray}
n_4=2 \alpha -1= 1.45+0.17 \ln \left(\ln  (\frac{Q^2}{Q_1^2})+1 \right).  
\end{eqnarray}
The result of both
the perturbative gluon solution and the hard gluon solution are given in 
Table 3 (middle columns). We note that
the quality of the fits is about the same as those obtained in Table 1.
The transverse terms prefer a soft Pomeron with a rather strong $Q^2$
dependence of the exponent $n_2$. We also observe that, for the hard
gluon solution, the value of the
parameter $x_0$ is quite different from the fit in table 1 (cf. the
footnote on p.2). We conclude that
both solutions are consistent with
the theoretical expectation that the longitudinal part of our parametrization
is hard, i.e. its energy dependence can be described by the gluon structure
function.

We have also tested the possibility of identifying the $x_{\PO}$ dependence 
of the longitudinal term with the structure function $F_2$ of the proton.
It has been shown ~\cite{F2} that the $F_2$ slope in $x$ can be parametrised 
as follows:
\begin{eqnarray}
\alpha=\frac{d \ln F_2}{d ln 1/x} = 1.15 + 0.09 \log_{10} Q^2,
\end{eqnarray}
where the scale of $Q^2$ is $1$ GeV$^2$.
The exponent $n_4$ can then be expected to vary as
\begin{eqnarray}
n_4=2 \alpha -1= 1.31+0.18 \log_{10} \frac{Q^2}{Q_1^2}  
\end{eqnarray}
In our fit we put the scale $Q_1^2=1$GeV$^2$. The results are shown in the righmost column
of Table 3. Again, the transverse terms in our parametrization 
want to have a soft Pomeron with a rather large $Q^2$ dependence. The
quality of the fits is about the same as that of the first fits in Table 1. 
Changing the scale to $Q_1^2=4$GeV$^2$ or treating it as a free parameter
does not change the result very much. Alltogether, the parameters of these
fits are very similar to the previous one with the gluon structure function.

\begin{table}[tb]
\vspace*{1.4cm}
\begin{scriptsize}
\begin{picture}(100,150)(0,0)
\put(5,80){\begin{tabular}{|c||c|c|c|} \hline
$\gamma$ & 7.36 $\pm$ 0.88 & 8.09 $\pm$ 1.05 & 8.08 $\pm$ 1.06 \\
\hline \hline
$\alpha_{R}$ & 0.40 $\pm$ 0.03 & 0.58 $\pm$ 0.09 & 0.59 $\pm$ 0.09 \\
$N$ & 38.7 $\pm$ 4.7 & 19.4 $\pm$ 11.0 & 18.3 $\pm$ 10.2 \\
\hline
$A$ & 0.40 $\pm$ 0.06 &  0.39 $\pm$ 0.08 & 0.30 $\pm$ 0.07 \\
$B$ & 0.17 $\pm$ 0.03 & 0.17 $\pm$ 0.04 & 0.14 $\pm$ 0.04 \\
$C$ & 0.048 $\pm$ 0.023 & 0.13 $\pm$ 0.03 & 0.14 $\pm$ 0.03 \\
\hline 
$n_2^0$ & 1.12 (fixed) & 1.00 $\pm$ 0.05 & 1.00 $\pm$ 0.06 \\
$n_2^1$ & 0. (fixed) & 0.26 $\pm$ 0.03 & 0.26 $\pm$ 0.03 \\
$n_4^0$ & 1.22 $\pm$ 0.21 & -   & -   \\
$n_4^1$ & 0.30 $\pm$ 0.19 & -   & -   \\
\hline 
$x_0$ & 0.19 $\pm$ 0.03 & 0.10 $\pm$ 0.04  & 0.12 $\pm$ 0.02 \\
\hline
$\chi^2$ & 254.2 & 190.9 & 186.7  
\\ \hline
\end{tabular}}
\put(240,80){\begin{tabular}{|c||c|c|c|} \hline
$\gamma$ & 0.70 $\pm$ 0.11 & 0.16 $\pm$ 0.11  & 0.23 $\pm$ 0.13 \\
\hline \hline
$\alpha_{R}$ & 0.40 $\pm$ 0.04 & 0.75 $\pm$ 0.03  & 0.76 $\pm$ 0.03 \\
$N$ & 38.6 $\pm$ 4.8 & 7.9 $\pm$ 1.5  & 7.3 $\pm$ 1.3 \\
\hline
$A$ & 0.086 $\pm$ 0.001 & 2.01 $\pm$ 0.51  & 1.16 $\pm$ 0.29 \\
$B$ & 0.045 $\pm$ 0.005 & 0.75 $\pm$ 0.21  & 0.37 $\pm$ 0.11 \\
$C$ & 0.018 $\pm$ 0.011 & 1.70 $\pm$ 0.67  & 1.28 $\pm$ 0.42 \\
\hline 
$n_2^0$ & 1.12 (fixed) & 1.04 $\pm$ 0.09  & 1.05 $\pm$ 0.06 \\
$n_2^1$ & 0. (fixed) & 0.30 $\pm$ 0.07  & 0.29 $\pm$ 0.05 \\
$n_4^0$ & 1.44 $\pm$ 0.06 & -   & -   \\
$n_4^1$ & 0.09 $\pm$ 0.09 & -   & -  \\
\hline 
$x_0$ & 0.41 $\pm$ 0.04 & 0.014 $\pm$ 0.002  & 0.022 $\pm$ 0.003  \\
\hline
$\chi^2$ & 291.2 & 210.0   & 208.3  
\\ \hline
\end{tabular}}
\end{picture}
\end{scriptsize}
\caption{Parameters obtained for the perturbative gluon (on the left) and hard gluon
(on the right) solutions of the
fit. The first fit is obtained by imposing the value of the soft pomeron
exponent for $n_2$ and in the second and third fit respectively, 
$n_4$ follows the gluon slope in $x$ and the $F_2$ slope
in $\log 1/x$ (cf text). }
\label{t1.0}
\end{table}

{\bf 5.} We conclude with a few general comments. Within the parametrization
suggested in ~\cite{bartels} have tried to clarify
the differences between the two possible solutions for the H1 94 data, 
and their relation to the
previously reported H1 solution with the hard gluon.
One of the two solutions (named `perturbative gluon`) is closer to what one
expects from a straightforward extrapolation of perturbative QCD calculations,
whereas the other one (called `hard gluon`) is closer to the previous H1 
result. The main difference lies in the $\beta$ dependence of the
$q\bar{q}g$ contribution.
Apart from a small difference in $\chi^2$ (in favor of the perturbative
gluon solution), we also found some evidence that, when we impose kinematic 
cuts on the data the hard gluon solution is slightly less stable 
than the perturbative gluon solution. 

As we have discussed in the 
beginning of this letter the parametrization for $F_2^D$ is based upon an
attempt to interpolate between the QCD parton model (hard region) and
Regge physics (soft Pomeron). In this first step of testing this approach
we have given much freedom to our parametrization. In a future step
it will be interesting to see how specific final states fit into this
parametrization and maybe used to constrain the parameters of our fit. 
Two obvious candidates are the production of longitudinal
vector mesons which should be contained in the third term of our 
parametrization, and the production of $q\bar{q}$ jets (contained in the 
first term) and $q\bar{q}g$ jets (inside the second term) from the transverse 
photon. The measured cross sections of these final states and a careful
analysis of their topological features could be used to
estimate the parameters of our ansatz. Moreover, their measurement 
can decide which of the two
solutions discussed in this letter is to be preferred.\\ \\   
{\bf Acknowledgement:} We thank J.Dainton for a careful reading of our 
manuscript and for his helpful criticism. We also thank R.Peschanski
for a careful reading of the manuscript.\\ \\
{\bf Note added in proof:} Another fit of our model to the combined 94 and the 
(preliminary) 95 low $Q^2$ data of H1  
has been presented at the Brussels DIS98 (Talk presented by T.Nicholls) 
conference. The extrapolation into 
the region of smaller $Q^2$ values works quite reasonably.
More recently (H1 Collab., Contribution to ICHEP98, Vancouver, July 98), 
our fit has been extrapolated also into the large $Q^2$ region
($200$ GeV$^2$ $<$ $Q^2$ $<$ $800$ GeV$^2$). We feel rather hesitant in 
using our simple parametrization in this region. First, as we have said 
before, extending the $Q^2$ region in such a dramatic way our parametrization
needs more terms and a more accurate way of handling the $Q^2$ evolution.
Secondly, in the large $Q^2$ region region of HERA $x_{\PO}$ is not really
small, and the influence of the secondary 
exchange term (4) of our parametrization starts to dominate.
Comparison of our fit with the large-$Q^2$ data,
therefore, does not only test the $Q^2$ evolution of our parametrisation
but also of the pion structure 
function used in the fit.

\eject

\section{References}


\end{document}